\documentclass[useAMS,usenatbib]{mn2e}
\usepackage{lscape,graphicx,natbib,longtable}

\title[Spectropolarimetry of the massive post-Red Supergiants IRC +10420 and HD 179821]{Spectropolarimetry of the massive post-Red Supergiants IRC +10420 and HD 179821}
\author[M. Patel et al.]{M. Patel$^{1,2}$, R.D. Oudmaijer$^{1}$, J.S. Vink$^{3,4}$,  J.E. Bjorkman$^{5}$, B. Davies$^{1,6}$, 
\vspace*{2mm} \\ {\LARGE \rm \noindent M.A.T. Groenewegen$^{7}$, A.S. Miroshnichenko$^{8}$, J.C. Mottram$^{1}$} \\
$^{1}$School of Physics $\&$ Astronomy, University of Leeds, Woodhouse Lane, Leeds LS2 9JT, UK \\
$^{2}$Astrophysics Group, Imperial College London, Blackett Laboratory, Prince Consort Road, London SW7 2AZ, UK \\ 
$^{3}$Armagh Observatory, College Hill, Armagh BT61 9DG, Northern Ireland, UK \\ 
$^{4}$Lennard-Jones Laboratory, Astrophysics, Keele University, Keele ST5 5BG, UK \\
$^{5}$Ritter Observatory, M.S. 113, Department of Physics and Astronomy, University of Toledo, Toledo, OH 43606-3390, USA \\
$^{6}$Chester F. Carlson Center for Imaging Science, Rochester Institute of Technology, 54 Lomb Memorial Drive, Rochester, NY 14623-5604, USA\\ 
$^{7}$Instituut voor Sterrenkunde, K.U.Leuven, Celestijnenlaan 200 D, 3001 Leuven, Belgium\\ 
$^{8}$Department of Physics and Astronomy, University of North Carolina at Greensboro, P.O. Box 26170, Greensboro NC 27402-6170, USA} 

\begin{document}
\date{}

\pagerange{\pageref{firstpage}--\pageref{lastpage}} \pubyear{2009}

\maketitle

\label{first page}

\begin{abstract}

We present medium resolution spectropolarimetry and long term
photo-polarimetry of two massive post-red supergiants, IRC~+10420 and
HD~179821.  The data provide new information on their circumstellar
material as well as their evolution. In IRC~+10420, the polarization
of the H$\alpha$ line is different to that of the continuum, which
indicates that the electron-scattering region is not spherically
symmetric.  The observed long term changes in the polarimetry can be
associated with an axi-symmetric structure, along the short axis of
the extended reflection nebulosity. Long term photometry reveals that
the star increased in temperature until the mid-nineties, after which
the photospheric flux in the optical levelled off. As the photometric
changes are mostly probed in the red, they do not trace high stellar
temperatures sensitively. And so, it is not obvious whether the star
has halted its increase in temperature or not. For HD~179821 we find
no polarization effects across any absorption or emission lines, but
observe very large polarization changes of order 5\% over 15
years. During the same period, the optical photometry displayed modest
variability at the 0.2 magnitude level. This is unexpected, because
large polarization changes are generally accompanied by strong
photometric changes. Several explanations for this puzzling fact are
discussed. Most of which, involving asymmetries in the circumstellar
material, seem to fail as there is no evidence for the presence
of hot, dusty material close to the star.  A caveat is that the
sparsely available near-infrared photometry could have missed periods
of strong polarization activity.  Alternatively, the variations can be
explained by the presence of a non-radially pulsating photosphere.
Changes in the photometry hint at an increase in temperature
corresponding to a change through two spectral subclasses over the
past ten years.

\end{abstract}

 
\section{Introduction}

Due to the steepness of the Initial Mass Function, massive stars
($\ga$ 8 M$_{\odot}$) are extremely rare. This is exacerbated by their
comparatively short lifetimes. Yet, although rare, these objects have
a crucial impact on the interstellar medium due to their strong winds
and high mass-loss rates, and can dominate the light output of entire
galaxies.

An area of current interest is that massive evolved stars are often
surrounded by bi-polar nebulae \citep{weis:2003}. Later in the
evolution of a star, using spectropolarimetry, it is now established
that the ejecta of supernovae deviate from spherical symmetry
(e.g. \citealt{Wang_etal:2003,Leonard_etal:2005}).  It is as yet
unclear whether this is due to asymmetric explosions, axi-symmetric
stellar winds or a pre-existing density contrast in the surrounding
material
(e.g. \citealt{Dwarkadas_Balick:1998,Dwarkadas_Owocki:2002}). Wind-axisymmetry
may imply fast rotation, and be related to the beaming of SN
explosions, which may be the origin of the extremely luminous, beamed
gamma-ray bursts (e.g. \citealt{Meszaros:2003,Mazzali_etal:2003}).

Here, we address the issue by investigating the circumstellar ejecta
of two yellow hypergiants, IRC~+10420 and HD~179821. These objects are
thought to have evolved off the post-Red Supergiant branch and are
still surrounded by mass ejected during a previous mass losing
phase. Only a few yellow hypergiants are known (see the review by
\citealt{de_Jager:1998}), and the number of such hypergiants with
circumstellar dust is even smaller - only IRC~+10420 and HD~179821
belong to this class (see for example
\citealt{Oudmaijer08}). Therefore, the study of these two unique
objects is important in its own right.

As these stars are distant (3-5 kpc), the direct imaging of their
innermost regions is currently beyond the reaches of current
technology, although interferometry is starting to resolve the winds
of evolved stars \citep{dewit_etal:2008}.  Observing with
spectropolarimetry allows us to probe regions much closer to the star
still.  Spectropolarimetry was first effectively used in the study of
classical Be stars using the presence of `line effects'
\citep{Poeckert:1975, Poeckert_Marlborough:1976}. These are changes in
polarization across spectral lines that have an emission
component. They occur because emission-line photons arise over a
larger volume than the stellar continuum photons. Consequently, the
emission-line photons undergo fewer scatterings as they `see' fewer
electrons, resulting in a lower polarization than the continuum. We
normally only observe a net polarization change if the geometry of the
electron scattering region is aspherical. Many authors have confirmed
that this technique provides evidence that envelope geometries around
Be stars are indeed disk-like \citep{dougherty_1992,
Quirrenbach_etal:1997, Wood_Bjorkman_Bjorkman:1997}.  More recently
the technique has been used to investigate the geometry of
circumstellar material around Herbig Ae/Be stars. Studies by
\cite{Oudmaijer_Drew:1999} and \cite{Vink_etal:2002} 
show that most Herbig stars exhibit line
effects, indicating aspherical electron-scattering regions. For a
recent review, see \cite{Oudmaijer:2007}.  With regard to evolved
stars, Davies et al. (2005) conducted a study of Luminous Blue
Variables (LBVs) using spectropolarimetry. They found that 50$\%$ of
the objects observed exhibited polarization changes across H$\alpha$,
indicating that some asphericity lies at the base of the stellar wind.
Furthermore, they found several objects for which the position angle
varied randomly with time, leading them to conclude that the wind
around these stars is clumpy (see also \citealt{Nordsieck_etal:2001}).

IRC~+10420 is now well accepted as a massive, evolved object
(e.g. \citealt{Jones_etal:1993, Oudmaijer_etal:1996} and
\citealt{Humphreys_etal:2002}). This is mainly based on its large
distance, high outflow velocity (40 kms$^{-1}$) and high luminosity
implied from the hypergiant spectrum.  The situation for HD~179821 is
less certain. The presence of non-radial pulsations coupled with
comparatively modest photometric changes suggest a massive nature
\citep{LeCoroller_etal:2003}. Furthermore, its circumstellar material
has a large expansion velocity of 30 km s$^{-1}$, as measured in CO,
suggesting the star is a supergiant \citep{Kastner_Weintraub:1995}. On
the other hand, the overabundance of s-process elements and the low
metallicity suggest HD~179821 is perhaps a lower mass post-AGB star
\citep{Zacs_etal:1996,Reddy_Hrivnak:1999,Thevenin_etal:2000}.

Although the latter's nature is a bit more uncertain, there are some
striking similarities between IRC~+10420 and HD~179821.  When observed
as part of a larger sample of post-AGB stars, these two objects are
often markedly different from the rest. In particular, their high
outflow velocities (the average outflow velocity for post-AGB and AGB
stars is 15 kms$^{-1}$) require much higher luminosities if powered by
radiation pressure alone \citep{Habing_etal:1994}. They were the only
objects that showed extensive reflection nebulae in a large survey by
\cite{Kastner_Weintraub:1995}. \cite{Jura_etal:2001} point out the
enormous difference between the space velocities of IRC~+10420 and HD
179821 when compared against low mass post-AGB stars. Furthermore,
both objects have an exceptionally strong O{\sc i} 7774 triplet
absorption feature indicating a high luminosity
(\citealt{Humphreys_etal:1973} and \citealt{Reddy_Hrivnak:1999}, based
on \citealt{slowik_1995}). We therefore proceed with both objects and
assume they are evolved post-Red Supergiants.

Recently, both objects have been observed at arcsecond resolution in
CO by \cite{Castro-Carrizo_etal:2007} who found, in accordance with
previous estimates, that their mass loss rates exceeded 10$^{-4} \,
\rm M_{\odot} yr^{-1}$ when they were in the RSG phase. The envelopes
show mild deviations from spherical symmetry in their data.

This paper is organised as follows. In Section 2, we review the
experimental setup and explain how the data has been reduced. We
present our results for each object in Section 3, and use the new
spectropolarimetric data together with past polarization measurements
to investigate the nature of the circumstellar environments around
each of the stars, which is discussed in Section 4.  We conclude in Section 5.

\begin {table*}
\centering
\begin {tabular}{llcccccccccc}
\hline
\centering
Object & Telescope & Date     & Julian Date & $\%$P & P.A. (Deg)  & $\%$P (H$\alpha$) & P.A. (H$\alpha$) (Deg) \\
\hline
  HD~179821 \\
            & AAT  & 15/09/02 & 2452533  & 1.99 $\pm$ 0.11 & 40 $\pm$ 2 \\
            & WHT  & 30/09/04 & 2453279  & 2.00 $\pm$ 0.15 & 36 $\pm$ 3 \\
		 					 
  IRC~+10420 \\
            & NOT  & 16/01/98 & 2450830  & 1.95 $\pm$ 0.05 & 174 $\pm$ 1 \\
            & NOT  & 16/05/98 & 2450950  & 1.80 $\pm$ 0.03 & 174 $\pm$ 1 \\
            & AAT  & 18/09/02 & 2452536  & 2.12 $\pm$ 0.11 & 173 $\pm$ 2 & 1.28 $\pm$ 0.11 & 10 $\pm$ 2\\
            & AAT  & 15/08/03 & 2452867  & 2.35 $\pm$ 0.11 & 179 $\pm$ 1 & 1.61 $\pm$ 0.11 & 11 $\pm$ 1\\
            & WHT  & 30/09/04 & 2453279  & 3.40 $\pm$ 0.15 & 174 $\pm$ 2 & 1.93 $\pm$ 0.15 & 11 $\pm$ 2\\
\hline
\end{tabular}
\caption{New polarimetric observations of HD~179821 and IRC~+10420
measured in the $R$ band. For the spectropolarimetric data (those
taken at the AAT and WHT), the polarization was measured in the
continuum region close to H$\alpha$, while the final columns are the
polarizations at the line centers. The H$\alpha$ line-center
polarization of HD~179821 was not calculated as no line effect was
observed, and therefore this measurement would be equal to the
continuum polarization. The systematic error of the AAT and WHT
spectropolarimetric data is estimated to be of order 0.10\% and
0.15\%, respectively.  }
\label{T:SUMDATA} 
\end {table*}

\begin {table*}
\centering
\begin {tabular}{llrrrrrrrr}
\hline
Julian Date & Telescope  & $B-V$&   $V$& $V-R$& $R-I$  &  $J$ &  $H$ &  $K$  \\ 
\hline
2450268     &  CST       &      &      &      &      &    5.62&      &       \\ 
2450303     &  CST       &      &      &      &      &    5.36&  4.40&  3.45 \\ 
2450643     &  CST       &      &      &      &      &    5.40&  4.44&  3.50 \\ 
2450690     &  CST       &      &      &      &      &    5.40&  4.45&  3.49 \\ 
2450707     &  CST       &      &      &      &      &    5.37&  4.43&  3.38 \\ 
2450950     &  NOT       & 2.76 & 11.06&  2.40&  1.70&        &      &       \\ 
2450985     &  CST       &      &      &      &      &    5.38&  4.44&  3.48 \\ 
2451037     &  TSAO      & 2.58 & 11.12&  2.42&  1.60&        &      &       \\ %
2451039     &  TSAO      & 2.67 & 11.11&  2.41&  1.58&        &      &       \\ 
2451042     &  TSAO      & 2.58 & 11.13&  2.42&  1.54&        &      &       \\ 
2451043     &  TSAO      & 2.73 & 11.06&  2.49&  1.61&        &      &       \\ 
2451047     &  TSAO      & 2.51 & 11.01&  2.41&  1.58&    5.63&  4.55&  3.73 \\ 
2451048     &  TSAO      & 2.75 & 11.03&  2.47&  1.60&    5.43&  4.57&  3.58 \\ 
2451050     &  TSAO      & 2.67 & 11.01&  2.46&  1.66&        &      &       \\ 
2451052     &  TSAO      & 2.75 & 11.03&  2.47&  1.59&    5.35&  4.44&  3.53 \\ 
2451057     &  TSAO      & 2.76 & 10.96&  2.47&  1.64&        &      &       \\ 
2451063     &  TSAO      & 2.60 & 10.95&  2.43&  1.59&        &      &       \\ 
2451075     &  TSAO      & 2.70 & 11.16&  2.52&  1.60&        &      &       \\ 
2451082     &  TSAO      & 2.66 & 11.09&  2.47&  1.59&        &      &       \\ 
2451083     &  TSAO      & 2.72 & 11.05&  2.48&  1.60&        &      &       \\ 
2451087     &  CST       &      &      &      &      &    5.24&  4.29&  3.33 \\ 
2451099     &  TSAO      & 2.62 & 11.16&  2.56&  1.66&        &      &       \\ 
2451100     &  TSAO      & 2.71 & 11.16&  2.54&  1.66&        &      &       \\ 
2451103     &  TSAO      & 2.76 & 11.13&  2.51&  1.54&        &      &       \\ 
2451104     &  TSAO      & 2.72 & 11.03&  2.48&  1.59&        &      &       \\ 
2451147     &  CST       &      &      &      &      &    5.35&  4.42&  3.46 \\ 
2451245     &  TSAO      &      &      &      &      &    5.32&  4.47&  3.56 \\ 
2451292     &  CST       &      &      &      &      &    5.40&  4.45&  3.57 \\ 
\hline
\end{tabular}
\caption{ Photometry of IRC~+10420. The data come from the Carlos
S\'{a}nchez Telescope (CST), the Nordic Optical Telescope (NOT) and
the Tien-Shan Astronomical Observatory (TSAO), see text for details.
Typically,  the photometric errors are of order 0.01-0.03 magnitude.}
\label{T:PHOTDATA} 
\end {table*}

\section{Observations}

We describe our most recent data in detail. Similar observations taken
earlier are briefly discussed, with relevant differences highlighted.
The linear spectropolarimetric data were taken on the night of 30
September 2004 using the ISIS spectrograph on the 4.2m William
Herschel Telescope (WHT), La Palma.  A MARCONI2 CCD detector with a
R1200R grating was used. This yielded a spectral coverage of 6150-6815
\AA$ $ and gave a spectral resolution of 34 kms$^{-1}$ at
H$\alpha$. The seeing was about $2''$, and a slit with a width of 1
arcsec was used.  The linear polarimetric component of the data was
analysed using the polarization optics equipment present on the ISIS
spectrograph. The object and the sky background were simultaneously
observed using additional holes in the dekker mask. A calcite block
was then used to split the rays into two perpendicularly polarized
beams (the o and e rays). A complete data set therefore, consists of
four spectra observed at 0$^\circ$ and 45$^\circ$ (to measure Stokes
Q) and 22.5$^\circ$ and 67.5$^\circ$ (to measure Stokes U).

All data reduction steps used the {\sc figaro} software maintained by
Starlink and included bias subtraction, cosmic ray removal, bad pixel
correction, spectrum straightening and flat fielding. Wavelength
calibration was carried out using observations of a Copper-Argon lamp
taken throughout the run. The data were then imported into the package
{\sc ccd2pol} (also maintained by Starlink), to produce the Stokes Q
and U parameters. The degree of polarization and its PA can then be
found from
\begin{equation}
p^{2} = Q^{2} + U^{2}
\end{equation}

\begin{equation}
 \theta = 0.5\, \times \, \arctan\,(\frac{U}{Q})
\end{equation}

Finally, the data was position angle-calibrated using a set of
polarization standard stars observed during the run.

Polarimetric data should only be limited by photon statistics. For
example we expect an error of about 0.1$\%$ for a detection of 10$^6$
photons, and significantly smaller when large parts of the spectra are
binned. However, the presence of systematic errors such as scattered
light and instrumental polarization often outweigh these
photon-noise errors and we estimate the errors in the present data to
be of order 0.15\%. The observations of two zero-polarization objects
show that the instrumental polarization is less than 0.15$\%$.  We
have made no attempt to correct for either instrumental or
interstellar polarization (ISP), as these only add a constant (Q,U)
vector to the data and therefore will not affect whether we observe a
line effect.

\begin{figure*}
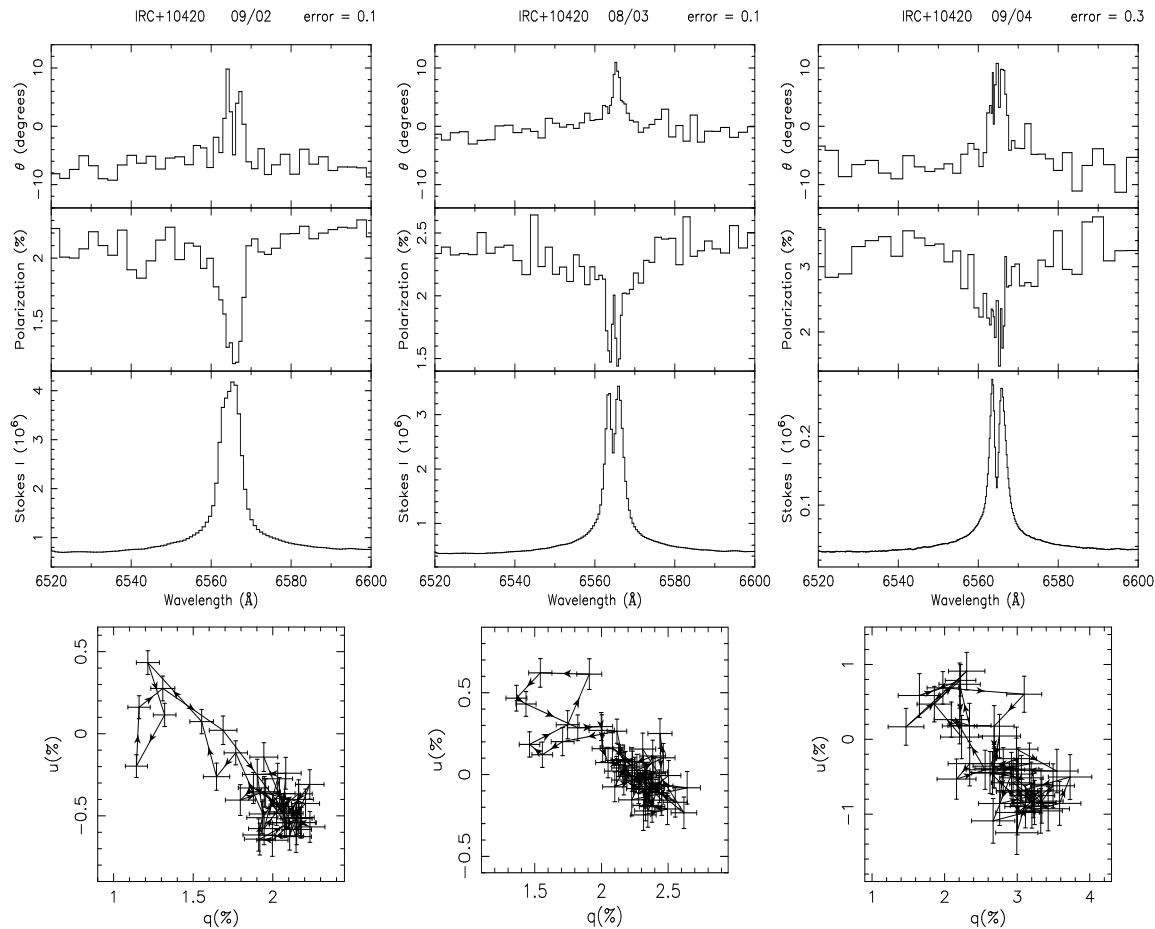

\centering
\includegraphics[width=50mm, height=80mm]{./specpolpaper_irc10420_2002ha.ps}
\includegraphics[width=50mm, height=80mm]{./specpolpaper_irc10420_2003ha.ps}
\includegraphics[width=50mm, height=80mm]{./specpolpaper_irc10420_2004ha.ps}
\vfill
\vspace{2mm}
\hspace{4mm}
\includegraphics[width=40mm, height=40mm]{./specpolpaper_irc10420_2002haQU.ps}
\hspace{9mm}
\includegraphics[width=40mm, height=40mm]{./specpolpaper_irc10420_2003haQU.ps}
\hspace{9mm}
\includegraphics[width=40mm, height=40mm]{./specpolpaper_irc10420_2004haQU.ps}
\hspace{5mm}
\caption{Spectropolarimetry of IRC~+10420 around H$\alpha$ at
different epochs. 
The upper diagrams are `triplots' of the spectropolarimetric data,
which were taken one year apart from one another. The bottom panel of
the triplot presents the direct intensity spectrum, the data normally
obtained from spectroscopy. The middle panel shows the polarization
measured as a function of wavelength. The upper panel shows the
corresponding polarization angle.  The data are adaptively binned such
that each bin has the same observational error derived from Poisson
statistics. Here we have chosen errors of 0.1\%, 0.1\% and 0.3\%
respectively for the 2002, 2003, and 2004 data. Below the triplots,
the Stokes Q,U vectors are plotted against each other, and binned to
the same precision as the spectra.  Depolarization around H$\alpha$ is
seen each time and the associated excursions in Q-U space all point in
the same direction.}
\label{F:IRC10420_ha}
\end{figure*}

\subsection{Additional Data}

\subsubsection{Spectropolarimetry}

Further data of IRC~+10420 and HD~179821 was obtained during previous
observing runs.  Spectropolarimetric observations of IRC~+10420 were
obtained using the 3.9m Anglo-Australian Telescope (AAT) on 18/9/2002
and 15/8/2003. HD~179821 was observed on 15/9/2002 at the AAT.  The
2002 data are taken with a lower spectral resolution (2.1 $\rm \AA$)
and the 2003 data are taken at comparable resolution to the current
WHT data.  We refer to \cite{Davies_etal:2005} for more details on the
observations taken at the AAT in 2002 and 2003.

\subsubsection{Photometry and polarimetry}

We also obtained additional photometric and polarimetric data of IRC
+10420.  Hitherto unpublished broad band polarimetry was obtained in
January and May 1998 using the 2.5m Nordic Optical Telescope in La
Palma, and on one of these nights {\it BVRI} photometry were obtained.
These observations are described in \cite{oud_2001}.  Johnson {\it
BVRIJHK} observations were obtained in 1998 on the 1 m telescope at
the Tien-Shan Astronomical Observatory (Kazakhstan) with the
2d-channel photometer-polarimeter FP3U of the Pulkovo Observatory
\citep{Bergner_etal:1988}.  Further near-infrared photometry was
obtained employing the 1.5m Carlos S\'{a}nchez Telescope in Tenerife.
Details of the observing procedures can be found in
\cite{Kerschbaum_etal:2006}. The results are presented in
Tables~\ref{T:SUMDATA} and ~\ref{T:PHOTDATA}.   {\it R} band
polarimetric data and optical and near-IR photometric data taken from
the literature, for both objects, is described in Tables
$\ref{T:LITDATA}$ and $\ref{T:LITDATAP}$ respectively.

\begin {table*}
\centering
\begin {tabular}{lcrlll}
\hline
\centering
Object & Date & {\it N} & Source   &  Method\\
\hline
  HD~179821 \\
& May 1989 - May 1991      & 4  & Parthasarathy et al. (2005) & Broadband \\
& October 1991             & 1  & Trammell et al. (1994)      & Spectropolarimetry\\
& August 1993              & 1  & HPOL database, Johnson (priv. communication) & Spectropolarimetry \\
& June 1997 - October 1998 & 30 & Melikian et al. (2000)      & Broadband \\
    	       	 	      			 
  IRC~+10420 \\

& May 1976 - June 1976     & 1  & Craine et al. (1976)        & Broadband \\
& May 1989                 & 1 & Johnson $\&$ Jones (1991)    & Broadband\\
& October 1991             & 1 & Jones et al. (1993)          & Broadband\\
& October 1991             & 1 & Trammell et al. (1994)       & Spectropolarimetry \\
\hline
\end{tabular}
\caption{$R$ band polarimetric observations of HD~179821 and IRC~+10420 taken from the literature. The observational period is given in the second column. The last three columns show the number of observations taken, the literature source of the data and the method. Although Craine et al. (1976) made several observations of IRC~+10420, these values were averaged to a single data point in our study. We measured the continuum close to H$\alpha$ from the spectropolarimetric data in order to facilitate  comparison with the {\it R} band data.}
\label{T:LITDATA} 
\end {table*}

\begin {table*}
\centering
\begin {tabular}{lccrl}
\hline
\centering
Object & Date & Band & N & Source \\
\hline
  HD~179821 \\
& March 1988 - April 1988 & {\it V,J,K} & 1,2 & Hrivnak et al. (1993) \\
& May 1990 - October 1999 & {\it V} & 169 & Arkhipova et al. (2001) \\
& May 1990 - November 2005 & {\it V} & 254 & ASAS catalogue  \\
& June 1992 & {\it J,K} & 1 & Kastner $\&$ Weintraub (1995)  \\
& October 1998 & {\it J,K} & 1 & 2MASS database  \\
    	       	 	      			 
  IRC~+10420 \\

& May 1996 - June 1996 & {\it J,K} & 2,2  & Humphreys et al. (1997) \\
& April 2001 & {\it J,K} & 4,5  & Kimeswenger et al. (2004) \\
& April 2002 - October 2004 & {\it V} & 104 & ASAS Catalogue \\
\hline
\end{tabular}
\caption{A summary of recent photometric observations of IRC~+10420 and HD~179821. In the fourth column we give the number of observations taken in each passband, respectively. For IRC~+10420, this acts as an update to the previous tables compiled by Jones et al. (1993) and Oudmaijer et al. (1996).}
\label{T:LITDATAP} 
\end {table*}



\section{Results }

\subsection{IRC~+10420}

\subsubsection{Spectropolarimetry}

In Fig. $\ref{F:IRC10420_ha}$ we have plotted the polarimetric data
around H$\alpha$ taken in 2002, 2003 and 2004.  These data show
`triplots' of the spectropolarimetry, along with graphs plotting the
Stokes Q,U vectors against each other.

\begin{figure*}
\centering
\includegraphics[width=80mm]{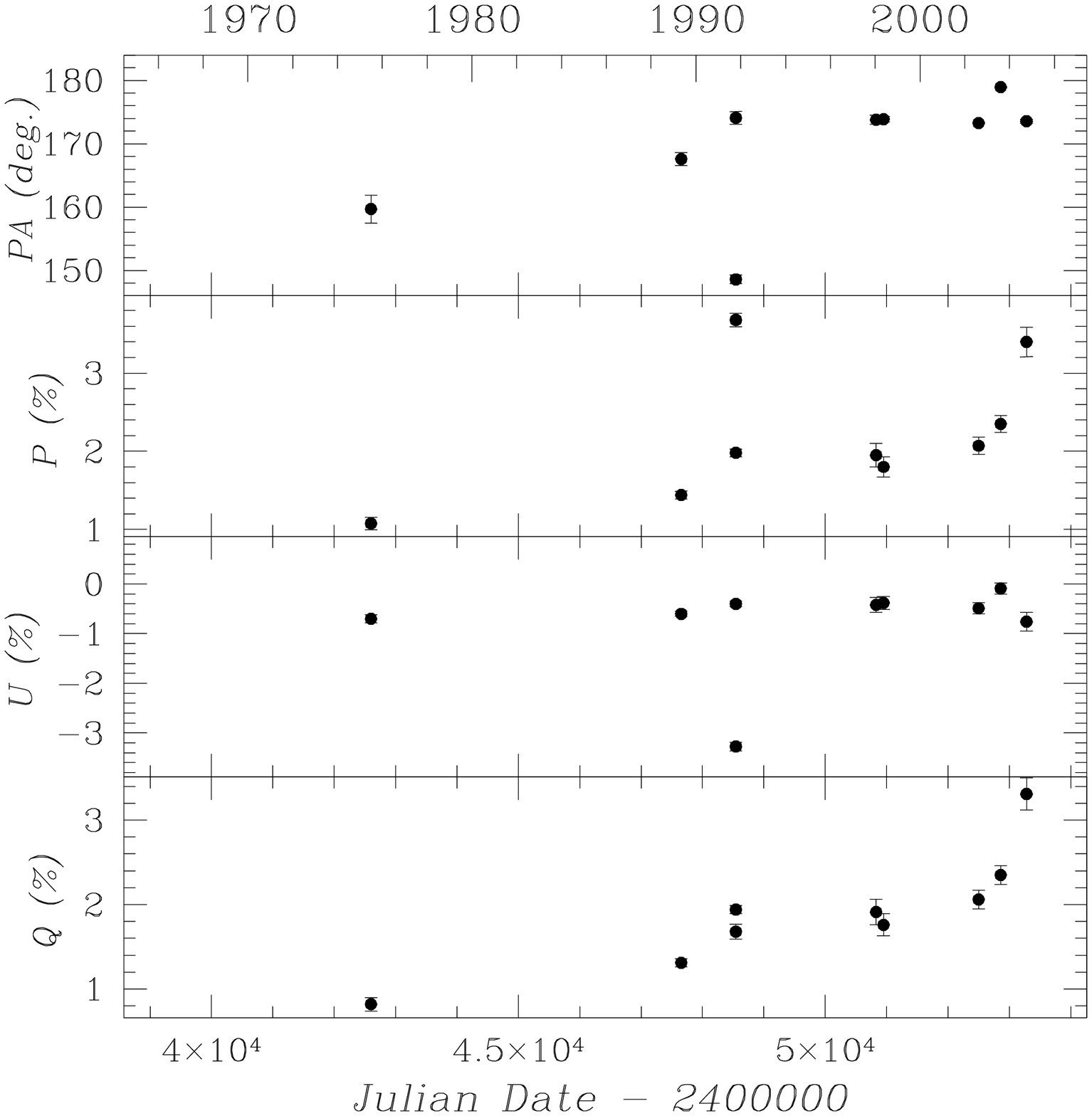}
\includegraphics[width=80mm]{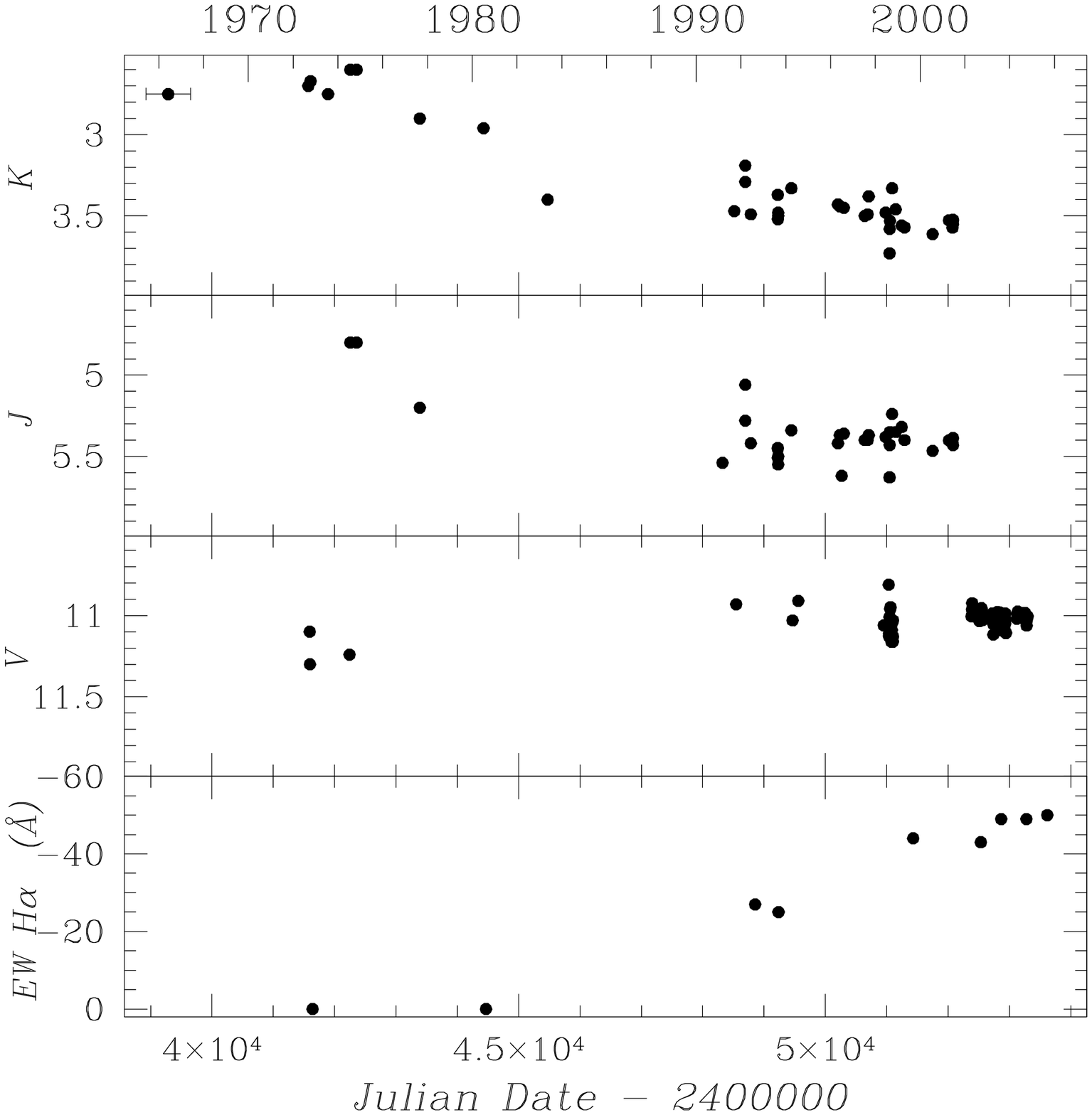}
\caption{Left : The polarization and position angle variability of
IRC~+10420 in the {\it R}-band. Both the polarization and PA show an
increase with time. The bottom 2 panels show the evolution of the
Stokes {\it Q} and {\it U} vectors. Right : An updated version from
the photometric variations presented in Oudmaijer et al
(1996). The change in the {\it J} band, which traces the stellar
photosphere, indicates a temperature increase of the star. Also
included is the evolution of the H$\alpha$ Equivalent Width which
displays a marked strengthening over the years.}
\label{F:IRC10420_var}
\end{figure*}

\begin{figure}
\centering
\includegraphics[width=70mm, height=70mm]{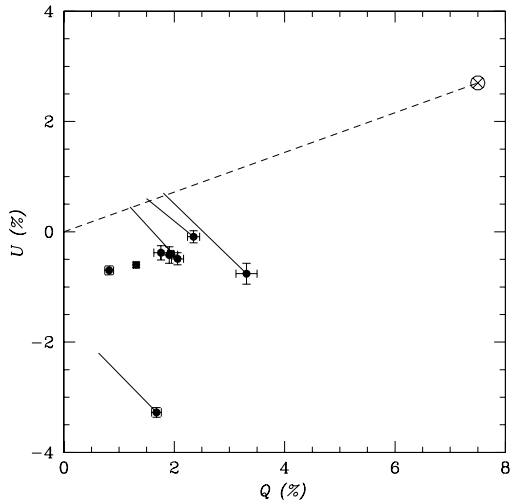}
\caption{ A Q-U diagram of IRC~+10420. The {\it R}-band (Q,U)
data-points come from the literature and our new broadband
polarimetry. The dashed line represents the vector towards the ISP,
denoted by the crossed circle, as derived by
\citet{Jones_etal:1993}. The continuum values from the
spectropolarimetric data are represented by dots and the straight
lines show the excursions to the line center polarization measured
from the spectropolarimetry.}
\label{F:IRC10420_varQU}
\end{figure}

IRC~+10420 displays H$\alpha$ emission with a total equivalent width
(EW) of --50 $\rm \AA$.  The emission line is unresolved in the lower
resolution 2002 data. The other, higher resolution, data indicate a
double-peaked and variable emission-line profile. In 2003, the red
peak is stronger than the blue peak, but in 2004 the blue peak has
become the stronger, while the central absorption component of the
double-peaked line has become deeper.  Comparison with published data
reveals that the red peak is very strong in the present spectrum.
\cite{Oudmaijer:1998} show data taken in 1993 of the object, where the
blue peak is 2 to 3 times stronger than the red one. The EW of the
line was --25$\rm \AA$, about half that observed now. This evolving
blue-to-red peak ratio was also noted by \cite{Klochkova_etal:2002}.

The spectropolarimetric data show large polarization changes across
the line. In each case, this line effect consists of a depolarization
over the central part of the line coupled with a rotation to larger
angles.  The associated Q-U plots (Fig. $\ref{F:IRC10420_ha}$) also
clearly show the line effects. The cluster of data points in the lower
right hand corners of the graphs are due to the stellar continuum,
while the excursions to the upper left hand corners trace the
polarization over the H$\alpha$ line. At first glance, the excursions
might seem to suggest the presence of structure. However, given that
the binning is at the 0.1\% and 0.3\% level respectively, and that the
`structure' is mostly due to single points offset by less than
3$\sigma$ from the global trends, we proceed under the assumption that
the excursions are linear.  The intrinsic polarization angle for IRC
+10420 is measured by the excursion from the line-center (which has a
lower contribution by the polarization due to electron scattering)
to the continuum. Using Eq. 2 this is 158$^{\rm o}$, with an estimated
error of 2$^{\rm o}$, and it does not seem to change with time.  The
magnitude of the excursion gives an indication of the strength of the
polarization change and is similar in 2002 and 2003, but almost twice
as large in 2004. Strictly speaking the length of the vector
represents a lower limit to the real depolarization because a finite
spectral resolution can wash out the line-effect. However, the 2002,
lower resolution data has a depolarization vector of similar length to
the higher-resolution, 2003 data. It is therefore plausible that the
depolarization (and thus the polarization of the continuum due to
electron scattering) is significantly stronger in the 2004 data.

The continuum polarization towards IRC~+10420 consists of three
components. Firstly, there is interstellar polarization due to
dichroic absorption of interstellar dust. This is constant with time
and is constant over a small part of the spectrum. In data such as in
Fig. $\ref{F:IRC10420_ha}$ this would add a constant vector to the
intrinsic Q,U spectrum. Importantly, the shape of the QU behaviour is
not affected by the ISP.  Secondly, some polarization may be due to
scattered light reflected off circumstellar dust grains. The presence
of scattered light is revealed by \cite{Kastner_Weintraub:1995} who
show that the polarization is significant at large distances from the
star. Later HST images show substantial extended emission in blue
light which is interpreted as scattered light as well
(\citealt{Humphreys_etal:1997}, \citeyear{Humphreys_etal:2002}; see
also \citealt{Davies_etal:2007}).  When obtaining spatially unresolved
data, such as here, any net polarization is due to the asymmetry of
the scattering material on the sky.  Thirdly, the continuum light can
be scattered off free electrons that are very close to the star. Here,
imaging polarimetry can not achieve the spatial resolution required to
study the electron scattering region. The line-effect however readily
reveals its presence.

\subsubsection{Long term photo-polarimetry}

Our new data are combined with literature data and are shown in
Figure~\ref{F:IRC10420_var}.  Over the past 30 years, both the
polarization and the angle have increased.  To illustrate this, the Q
and U Stokes vectors are also plotted.  The Stokes Q parameter
increases by about 2\%, while the U vector essentially remains
constant (see also Fig. $\ref{F:IRC10420_varQU}$).  Here we note the
\cite{Trammell_etal:1994} data. Their data point (3.7\% at 149$^{\rm
o}$) was taken in the same month as the \cite{Jones_etal:1993} data
which is 2.0\% at 174$^{\rm o}$. The Trammell et al. measurement
strongly deviates from the general, 30-year, trend in {\it P},
$\Theta$ and {\it U} and is also significantly different from data
taken in the same month. We will proceed assuming that the trend of
increasing polarization with time reflects the true long term
polarization behaviour of the object, and have a caveat that, perhaps,
the evolution is more complex. 
In Fig. $\ref{F:IRC10420_varQU}$ we plot the broad-band Stokes Q and U
vectors against one another. \cite{Jones_etal:1993} derive an
interstellar {\it R} band polarization of 8 per cent towards
IRC~+10420 - a value that is not inconsistent with the large reddening
towards the object. From their figures we measure their ISP to be
$Q_{R, \rm ISP} \approx 7.5\%, U_{R, \rm ISP} \approx 2.7\%$. This
value is indicated in
Fig. $\ref{F:IRC10420_varQU}$. Assuming the H$\alpha$ line center is
not polarized due to electron scattering, its net polarization is due
to a combination of ISP and scattering off the circumstellar dust.  We
can thus get a (very) rough estimate of the circumstellar dust
polarization by taking the difference between the H$\alpha$ line
center polarization and the ISP value.  Our observed line center
polarizations are close each other. Remarkably, the recent H$\alpha$
line polarization values all lie on the vector connecting the ISP with
the origin.  Their average distance to the ISP value is ($Q=-6.0\%,
U=-2.15\%$), corresponding to a circumstellar dust polarization of
($P=6.4\%, \Theta=9.9^{\rm o}$). The true value may have a larger
polarization at a smaller angle, as the polarization in the center of
the line is probably affected by the resolution.  Note that an
interesting situation has arisen where the circumstellar dust
polarization effectively cancels part of the ISP resulting in smaller
net polarization values.  Using Eq.~2, we measure an intrinsic
polarization angle for the variations of $6\pm 2^{\rm o}$. This is
very close to the orientation found for the circumstellar dust
polarization described above and pinpoints the changes in
circumstellar dust polarization as a likely cause for the observed
variations.

The photometric variability has been discussed extensively by
\cite{Jones_etal:1993} and \cite{Oudmaijer_etal:1996}. Here we give an
update on the long term variability (see Table $\ref{T:LITDATAP}$).
The $K$ band data show a steady fading starting in the mid-seventies.
This dimming is mimicked by the data in the {\it J} band, which is
known to trace the stellar photosphere
\citep{Oudmaijer_etal:1996}. After around 1995, the {\it J} band
magnitudes appear to reach a steady state.

\begin{figure*}
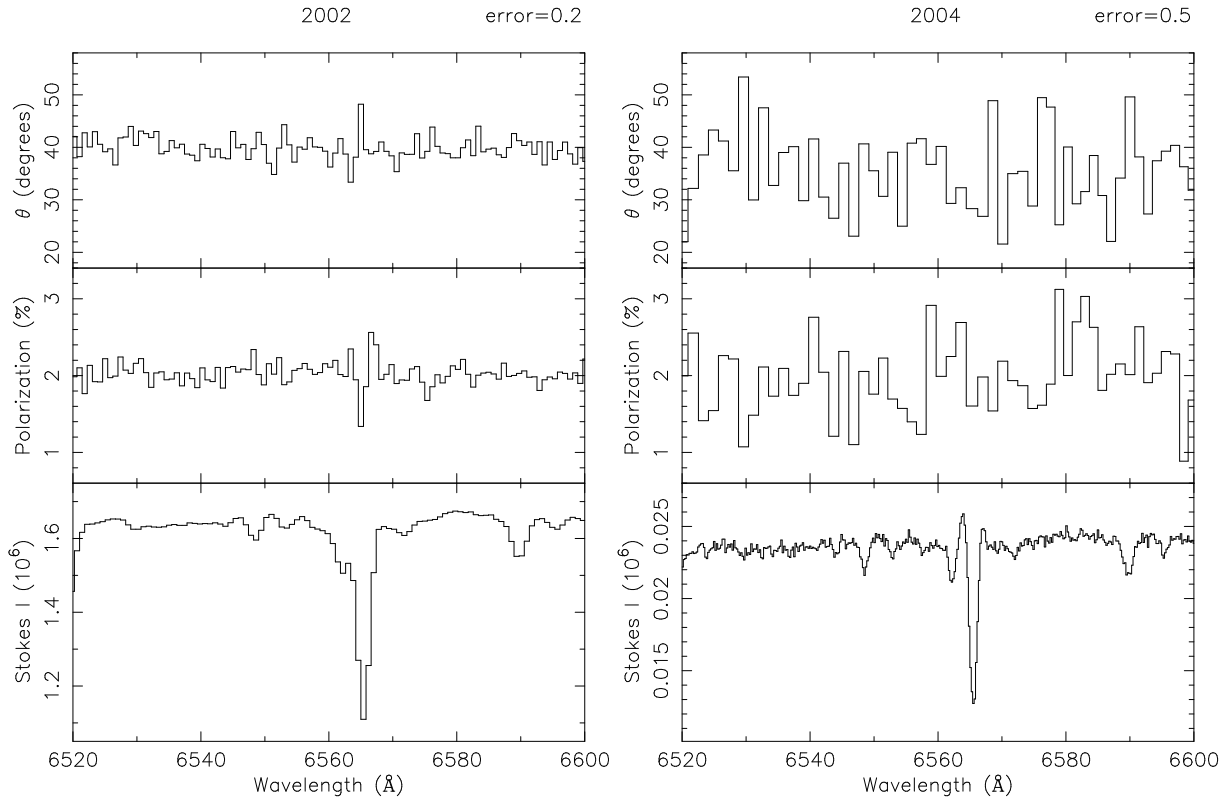

\centering
\includegraphics[width=80mm]{./specpolpaper_hd179821_ha_2002.ps}
\includegraphics[width=80mm]{./specpolpaper_hd179821_ha_2004.ps}
\caption{As in Fig.~\ref{F:IRC10420_ha}, but now for HD~179821 in 2002
(left) and 2004 (right). The data has been rebinned such that the
1$\sigma$ error in polarization corresponds to the value stated at the
top of the triplot, i.e. 0.2\% and 0.5$\%$ respectively, as
calculated from photon statistics.  }
\label{F:HD179821}
\end{figure*}

Long term changes have also been observed in the hydrogen
recombination lines. In 1992, \cite{Oudmaijer_etal:1994} discovered
NIR recombination emission lines that were previously seen in
absorption in 1984.  \cite{Irvine_Herbig:1986} detected H$\alpha$
emission in 1986, while previous observations did not show emission.
Therefore, the onset of hydrogen recombination emission can be traced
to a period between 1984 and 1986 \citep{Oudmaijer_etal:1994}. This
coincides precisely with a period of strong photometric changes.

EW data of H$\alpha$ were collected from the literature as well. It is
not trivial to obtain these data because of the large width of the
line. For example, EW measurements from high resolution echelle data
are not possible as the free spectral range is too small to properly
define a baseline and measure the equivalent width accurately.  We
therefore restrict ourselves to results from spectra with a large free
spectral range - essentially no echelle observations. Data are taken
from \cite{Oudmaijer_etal:1994}, \cite{Oudmaijer:1998},
\cite{Humphreys_etal:2002}, this paper and \cite{Davies_etal:2007}
respectively. The results are also shown in
Fig.~\ref{F:IRC10420_var}. H$\alpha$ has been increasing in strength,
going from upper limits of the EW in the seventies to --23 $\rm \AA$
in the early nineties to --50 $\rm \AA$ in 2005. As the optical
brightness has remained constant in that time, these EW changes
reflect true line flux changes and imply that more ionized material is
present.

\subsection{HD~179821}

\subsubsection{Spectropolarimetry}

Spectropolarimetric data of HD~179821 are shown in
Fig. $\ref{F:HD179821}$.  No polarization changes are visible across
the faint H$\alpha$ line. The 2002 data has a higher signal-to-noise,
but a weaker H$\alpha$ emission than in 2004.  There are few reports
of the H$\alpha$ line in the literature, but it would appear it is
variable in the sense that it varies between weak emission above the
continuum (i.e. 10-20\% emission above the continuum was observed by
\citealt{Zacs_etal:1996}) and weak emission that hardly fills in the
underlying absorption as observed here.

The weak emission implies that the number of free scattering electrons
is rather limited, which is not inconsistent with the low surface
temperature of the star (its spectral type is G4~0-Ia,
\citealt{keenan_1989}). Therefore any continuum polarization due to
electron scattering is bound to be low. In such a case, extremely high
SNR data are required to reveal a line depolarization, if it were
present at all.  Alternatively, the hydrogen recombination line
emission in HD~179821 could be due to photospheric shocks. The object
has been found to show periodic photometric variations of 140 and 200
days (see \citealt{LeCoroller_etal:2003}). These variations are most
likely due to pulsations, and it is common to
observe H$\alpha$ emission due to shocked layers in the photosphere of
cool, pulsating stars (e.g. \citealt{schmidt_2004} and references
therein).

\subsubsection{Long term photo-polarimetry}

\begin{figure}
\centering \includegraphics[width=80mm,
height=100mm]{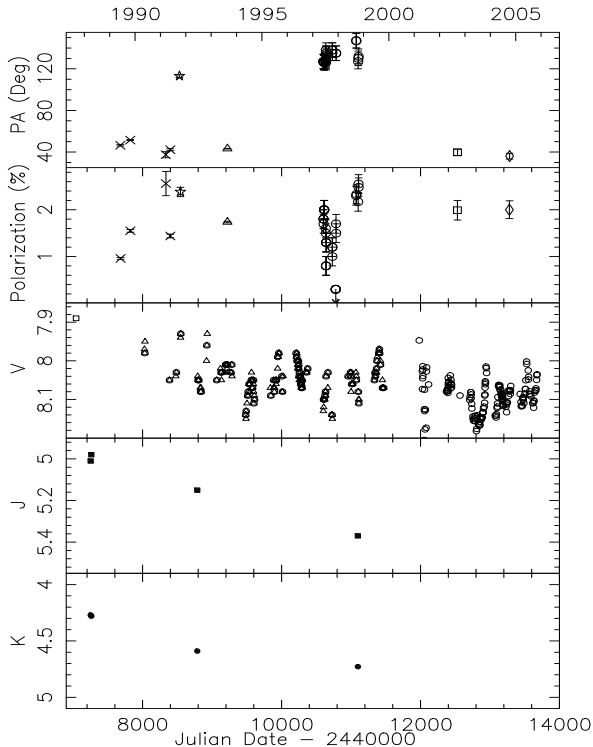}
\caption{The variability of HD~179821 with time. Plotted is the {\it
V}-band photometry provided by Hrivnak et al. (1989) [squares],
Arkhipova et al (2001) [triangles] and the ASAS Catalogue (Pojmanski,
2002) [circles]. The error associated with the photometry measurements
provided by Arkhipova is 0.005 mag and those from the ASAS Catalogue
are less than 0.06 mag. Plotted in the upper two plots are PA and
polarization data taken from Parthasarathy et al. (2005) [crosses],
HPOL database[triangle], Trammell et al. (1994) [star], Melikian et
al. (2000) [circles] and data presented in this paper from 2002
[squares] and 2004 [rhombi].  The PA exhibits large rotations over a
15 year period, while the polarization indicates the presence of
short-term variability. The lower panels show that HD~179821 has become fainter in $J$ and $K$.}
\label{F:HD179821_1}
\end{figure}

The photometry and polarimetry from this paper and the literature of
HD~179821 are plotted in Fig. $\ref{F:HD179821_1}$.  The longer term
optical variability is discussed by \cite{Arkhipova_etal:2001}.  From
1989 to about 1999 the star became fainter by about 0.1 magnitude, and
then returned to its original magnitude. \cite{Arkhipova_etal:2001}
suggest the photometry may be cyclic on long timescales. They also
find that the object is bluer in {\it U $-$ B} when the star is
brightest in {\it V}, which they attribute to pulsations. This is
confirmed by the detailed study on shorter term variations in the {\it
V} band presented by \cite{LeCoroller_etal:2003}, which are due to
pulsations with periods of order hundreds of days.  Here we report the
hitherto unnoticed large change in the {\it J} and {\it K}
bands. These became fainter by around 0.4 magnitude in the period 1988
- 2000.

Figure $\ref{F:HD179821_1}$ shows that HD179821 experiences
polarization changes (the star is even observed to exhibit no
polarization at all for four consecutive days in November 1997),
suggesting that the star is intrinsically polarized. The
polarization varies more or less randomly between 0 and 2.5 per cent
over the past 15 years, but we can identify two clear values for the
PA, a low value around 40$^{\rm o}$ and a high value of around
120$^{\rm o}$.  When plotting these data in {\it QU} space
(Fig. $\ref{F:HD179821_QU}$) we can see that the data points are
distributed around what we can loosely describe as a straight line
through the origin.

The move through the origin of the {\it QU} diagram is responsible for
the large change in observed polarization angle. This must have
happened at least five times: the first occurred between the 1989-1991
\cite{Parthasarathy_etal:2005} dataset (PA=40$^{\rm o}$) and the 1991
\cite{Trammell_etal:1994} data point (PA=120$^{\rm o}$) taken half a
year later, when the object moved from the upper half to the lower
half in the QU diagram. After this, a move occurred between 1991 and
the HPOL data taken in 1993 (Johnson private communication, using
HPOL, \citealt{Wolff_etal:1996}). The object returned back to the
lower half as observed by \cite{Melikian_etal:2000}, who also directly
detected such an excursion in 1997-1998 (the non-detections which lie,
by definition, at the origin). Finally a crossing happened after their
observing run, but before our present data was taken.  The move
through the origin does not necessarily reflect a real 90$^{\rm o}$
rotation, as the ISP may contribute significantly to the total Q,U
vectors observed towards the object.

It is hard to determine the ISP towards any object, but we can make an
educated guess.  Since $A_{V} \sim \, 2$ for HD~179821
\citep{Hrivnak_etal:1989} and one typically finds that the percentage
polarization is equal to the extinction in magnitudes
(e.g. \citealt{oud_2001}), we can expect an ISP of up to 2 per cent
towards the object. A value for the interstellar polarization
angle may be estimated from the polarization of surrounding field
stars. To this end we investigated the compilation of polarized stars
due to \citet{Heiles_2000}. The local ISP appears very ordered towards
this part of the sky. We selected all objects with a polarization
measurement larger than 0\% and within a radius of 5 degrees from the
position of HD 179821. These 22 objects have an average polarization
angle of 50$^{\rm o}$ with a scatter of 24$^{\rm o}$.  The catalogue
also lists the photometric distances, and the average angle is the
same for the 11 furthest ($>$500pc) objects. Whatever the magnitude of
the polarization, these values indicate that the ISP can be located in
the top quadrants of Fig.~\ref{F:HD179821_QU}.  When taking these
angles at face value, and assuming the ISP is oriented in the same
direction at large distances as locally, then the ISP towards HD
179821 is 50$^{\rm o}$ at 2\%. This is close to the present-day value
of HD 179821, and, if true, would suggest that HD 179821 is presently
in a relatively quiescent state, yet the polarization is still
variable by up to a few per cent.  The instances when the object occupies
the bottom of the {\it QU} diagram can then be identified as 
periods where the polarization of HD 179821 is particularly strong.
The optical photometry does not signal any such activity however, and
continues its gradual decline to fainter magnitudes.

In short, during the last 15 years, the observed polarization of
HD~179821 crossed the origin in the QU-diagram several times. The
changes occur along a straight line and suggest some preferred
orientation, for which we derive a very rough estimate for the
intrinsic angle of 38$^{\rm o}$. In addition, the polarization changes
by 4-5\% , while only modest changes in the optical photometry of less
than 0.2 magnitude occur. Near-infrared photometry is sparse, and
indicates a gradual faintening of the star. The most recent data point
obtained on 12 October 1998, which was a few weeks before the end of
\cite{Melikian_etal:2000}'s campaign, i.e. when HD 179821 was, if the
ISP value is correct, still in a high state of polarization.


\section{Discussion}

We have observed the only two post-Red Supergiants with large infrared
excesses using multi-epoch spectropolarimetry and supplemented these
data with long term photo-polarimetry. We find some striking
similarities between the two objects, but also large differences. We
will start with the latter.

\subsection{The ionized gas}

The spectropolarimetry reveals a strong line-effect in the data of IRC
+10420, but none for HD~179821.  The latter displays weak H$\alpha$
emission and is much cooler than IRC~+10420. It is therefore not
surrounded by much ionized material. As electron-scattering is the
prime mechanism to produce line-effects, it can then be easily
understood why there is no line-effect towards HD~179821.

As described earlier, the intrinsic polarization angle due to electron
scattering for IRC~+10420 is measured by the excursion from the
line-center, and is 158$^{\rm o}$.  This angle remained constant
within the errorbars during the our three observing epochs, whereas
the observed polarization increased by 1.5\%. Since the excursion
across the line in QU space became larger between 2003 and 2004, we
can attribute this increase in continuum polarization to more
scatterings of continuum photons by free electrons rather than dust
particles. 

The presence of the line-effect is strong evidence that the electron
scattering region has a geometry that deviates from spherical
symmetry. The constant intrinsic polarization angle indicates that the
structure has remained stable.  The nature of the H$\alpha$ emitting
region, and especially its geometry, has been the topic of much
debate, often centering on the question whether it is spherical or not
(see the review of the literature by
\citealt{Davies_etal:2007}). \cite{Davies_etal:2007} use integral
field spectroscopy to study reflected light off circumstellar dust of
the emission line and show that the H$\alpha$ line emitting region is
not spherically symmetric.  Here we confirm this result.

It is unclear however what the polarization angle corresponds
to. Assuming that the scattering is optically thin, the structure
responsible for the scattering is oriented perpendicular to this at an
angle of 68$^{\rm o}$.  A comparison with the larger scale structure
observed by the Hubble Space Telescope data which traces the
reflection nebulosity \citep{Humphreys_etal:1997,Humphreys_etal:2002}
is inconclusive. The angle does not correspond straightforwardly with
either the long or short axis.  The Hubble data reveal very clumpy
structures at sub-arcsec scales and it proves impossible to link the
even smaller scales probed by the electron scattering, with the clumps
or larger scale structures in the imaging.

Finally, the increased ionization in IRC~+10420 may be either
explained by an increase of mass close to the star, due to either an
infall or outflow or due to an increase in temperature.

\subsection{Temperature evolution}

The photometry shows some remarkable similarities between the two
objects.  Both stars have been subject to studies of their spectral
energy distribution (SED, \cite{Oudmaijer_etal:1996} and
\cite{Hrivnak_etal:1989} for IRC~+10420 and HD~179821
respectively). From fits to their SEDs, it is clear that the
near-infrared {\it J} band photometry samples their photospheres and
not the hot dust. In both cases, this {\it J} band magnitude has
become fainter over the past decades.  At the same time, their {\it V}
band magnitudes varied only weakly, if at all.

The photometric changes in {\it J}, in particular the decrease in {\it
V $-$ J}, can be explained as being due to bolometric correction
effects associated with an increase in stellar temperature.  For IRC
+10420, this is consistent with published spectra (see
\citealt{Oudmaijer:1998, Klochkova_etal:1997}).  This trend is less
obvious more recently. It is likely that the {\it J} band magnitude
has reached a plateau, but it is not apparent whether this implies that
the temperature increase has halted or not. The {\it V$-$J} colour is
not very sensitive to changes in spectral type earlier than A0, so an
ongoing temperature increase is not ruled out based on the photometry.

{\it J} band data of HD~179821 is much more sparse, but the trend is
clear. The object has become fainter in {\it J}.  If the change in
{\it V$-$J} colour is due to changes in effective temperature, then
the star has become hotter during this period, in the same way as for
IRC~+10420. For an $A_V$ of 2 \citep{Hrivnak_etal:1989}, and combining
the intrinsic optical colours for supergiants as provided by
\cite{str_kur} and intrinsic near-infrared colours due to
\cite{koorn_83}, we find that the star must have evolved to an earlier
spectral type by about one or two spectral subtypes.

There is some ambiguity concerning the star's temperature as derived
from chemical abundance modelling. For example
\cite{Thevenin_etal:2000} favour a lower temperature, consistent with
the G spectral type. This is 1000K less than \cite{Reddy_Hrivnak:1999}
and \cite{Zacs_etal:1996} determinations using data taken
earlier. Thus, it is thus hard to draw any conclusions on the
evolution of the stellar temperature.  The near-infrared photometric
changes may be seen as independent evidence that the star appears to
have been evolving towards higher temperatures. Alternatively, the
temperature changes may be due to the formation of a
pseudo-photosphere (Smith et al. 2004).

The main finding in this section is that both objects appear to be
evolving towards the blue in the HR diagram on human
timescales. Judging from the plateauing of the {\it J} band photometry, IRC
+10420 may have slowed down its temperature evolution, or even halted
altogether. It has been proposed that the object is hitting the yellow
void \citep{de_Jager:1998, dejager_2001,Humphreys_etal:2002} and/or
the red-edge of the bi-stability zone \citep{smith_vink2004}.  In this
case, further evolution is prevented unless a large part of its
(pseudo-)photosphere is shredded. It will be interesting to follow the
object to see whether a large outburst indeed will happen.

\subsection{Polarization Changes}

\begin{figure}
\centering
\includegraphics[scale=0.3,angle=-90]{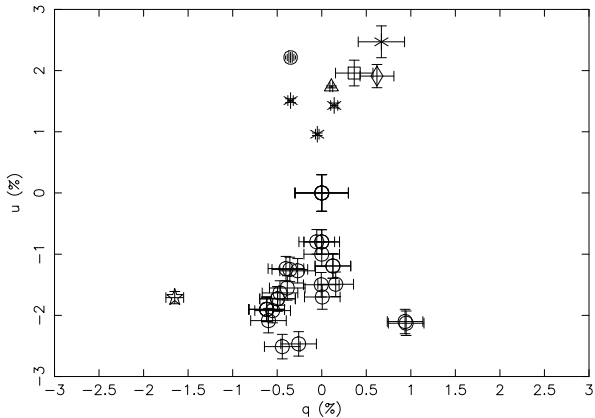}
\caption{A Q-U diagram of HD~179821. Each point corresponds to the
wavelength-averaged {\it R}-band continuum values of Q and U taken at
different epochs. There are  groupings of points in two separate areas
of the plot, one corresponding to low PA angles (top) and the other to
high angles (bottom). The filled circle shows the position of the
ISP (see text). The other symbols are as in the previous figure.
}
\label{F:HD179821_QU}
\end{figure}

A further similarity between the two objects is the significant
polarization variability over the past decades. IRC~+10420 increased
its polarization (as seen in the QU diagram,
Fig.~\ref{F:IRC10420_varQU}) by about 2.5\%.  HD~179821's polarization
changed by more than 5\%, following a straight line in QU space
(Fig.~\ref{F:HD179821_QU}). Such changes are not uncommon in stellar
objects, for example UX Ori variables and some T Tauri stars (see
e.g. \citealt{oud_2001, Grinin:1994}). For these objects, rotating
clumpy circumstellar disks can result in large polarization
changes. These occur when the clumps move into the line of sight of
the objects, reducing the optical brightness, often considerably by
many magnitudes, and thereby increasing the contribution of polarized
light.  However, here, in the case of both objects, the changes are
not accompanied by large changes in the optical photometry.   To
make matters worse, a direct consequence of the UX Ori scenario, to
explain the strong polarizations variations, is that the obscuring dust
is located close to the star in a compact, hot structure
\citep{Dullemond_etal:2003}. This means that a strong near-infrared
excess should be present and readily observable.  IRC~+10420 shows
evidence for a significant near-infrared excess due to hot dust. In
contrast, the near-infrared photometry of HD~179821 does not show any
sign of hot dust close to the star and is best explained by
photospheric emission \citep{Hrivnak_etal:1989}.

\subsubsection{The electron and dust scattering of IRC~+10420}

The, variable, continuum polarization of IRC +10420 is due to a
combination of dust and electron scattering.  The intrinsic PA
(10$^{\rm o}$) of the circumstellar dust polarization was derived
earlier from the vector connecting the ISP and the line centers of the
H$\alpha$ emission. This PA corresponds to the short axis observed in
the larger scale nebulosity of IRC +10420 and may signify a
circumstellar dusty disk structure.  The intrinsic angle of the
electron scattering region is 158$^{\rm o}$ and is not easily
identified with the larger scale structures observed in the HST
images. In the last observing epoch (2004), the line depolarization
vector was much stronger than previously observed. Although resolution
effects may have affected the amplitude of the excursion, the much
larger continuum polarization in 2004 indicates that enhanced electron
scattering is responsible for the most recent increase in
polarization. This is supported by the fact that the continuum
polarization moved along the line vector in the QU diagram
(Fig. ~\ref{F:IRC10420_varQU}) as well.  The PA measured for the QU
vector between the two recentmost dates is 166$\pm 3^{\rm o}$, very
close to the line polarization vectors.  Such an increase is not
inconsistent with the fact that the H$\alpha$ emission EW hardly
changed.  Polarization due to (optically thin) electron scattering is
more sensitive than (optically thick) H$\alpha$ emission for tracing
ionized material (see e.g.  \citealt{bjorkmank_2005}).

How does this compare to the longer term changes? The polarization
evolves along a more or less straight line with a slope corresponding
to a PA of 6$\pm 2^{\rm o}$, which closely corresponds to the
intrinsic PA of the dust scattering material. If the ISP determined
value by \citet{Jones_etal:1993} is correct, then it is fair to assume
that a decreasing circumstellar dust polarization is the main
contributor to the long term variability. A probable cause for this is
a clearing dust shell. This can be tested by producing SED model fits
to photometric data taken at several epochs.

In summary, by combining the intrinsic PA of both dust and electron
scattering and using the ISP, a relatively simple picture seems to
emerge from the variability of the polarization of IRC +10420. At
first a clearing dust shell dominates the polarization behaviour,
followed by a significant increase of polarization due to electron
scattering. The larger electron scattering is consistent with the
observed increase in H$\alpha$ emission, while the dust clearing can
be tested with modelling the circumstellar dust. 

This global picture is probably not the final answer however. Even the
data discussed here present more questions already.  For example, it
is not clear why the point taken by Trammell et al. in 1991 strongly
deviates from all other values (even one taken in the same month).
The H$\alpha$ depolarization they observe is consistent with what we
find ten years later, and confirms that part of the polarization is
due to electron scattering.  The large continuum value would indicate
a violent event that led to a polarization increase of several per
cent within weeks, however there is no other evidence to suggest such
sudden changes in other observations of IRC +10420. Similarly, it is
not obvious why the onset of H$\alpha$ line emission in the early
nineties, and its corresponding polarization of 1-1.5\%, is not
detected in the long term polarimetry, which follows a more or less
straight line from to the seventies to the current data. A speculation
is that a varying, rotating component of electron scattering was
present from the onset. Given the increase in H$\alpha$ emission, it
is possible that the electron scattering has become more optically
thick.  If part of the photons scatter more than once in their escape
from the electron scattering region, a full or partial flip in the
position angle can result (see e.g. \citealt{Vink_etal:2005}).

\subsubsection{But what about HD~179821?}

The changes observed for HD~179821 are less straightforward to
interpret.  The major question is how large changes over such an
extended period can occur while, apart from the polarization, other
observations of the star show mild variability at best. Firstly, we
note that the ISP is hardly likely to vary on human timescales. This
is supported by the results from a survey of polarization standard
stars which reveals no variability in the observed polarization down
to extremely low levels \citep{clarke:1994}. The H$\alpha$ emission is
insufficiently strong to suggest the presence of a large electron
scattering region, let alone one that it is variable enough to explain
the observed polarization changes. Below we discuss the several
possibilities that can account for the polarization and its
variability observed towards HD~179821.

\paragraph{Asymmetric circumstellar material} 

\smallskip

Having been forced to rule out electron scattering as being
responsible for the variability, we now consider scattering by
circumstellar dust. The presence of dust scattering was shown in the
imaging polarimetry by \cite{Kastner_Weintraub:1995}. The fact that
our, unresolved, data show a net polarization would normally indicate
that part of the scattering geometry deviates from spherical symmetry.
However, this is not supported by existing imaging data of the
circumstellar material of the object.  The scattering dust traced by
\cite{Kastner_Weintraub:1995} stretches over several arcseconds, but
appears to be distributed spherically symmetrically. HST images
published by \cite{Ueta_etal:2000} show small scale structure but
largely a round appearance.  
\citeauthor{Bujarrabal_etal:1992}'s \citeyear{Bujarrabal_etal:1992} CO
rotational data maps show a slightly elongated structure at larger
($\sim$5 arcsec) distances from the star at a PA of 50$^{\rm o}$.


Even if the slight asymmetry would be responsible for the observed net
polarization, it is hard to reconcile this with the observed
variability.  With a kinematic age of order thousands of years
(e.g. \citealt{Kastner_Weintraub:1995}), one would not expect changes
at a fraction of this period, i.e. 15 years, to still be traceable.  A
more practical objection to the extended emission being variable is
that the intensity of the scattered light rapidly drops as a function
of distance to the star, so there is simply not enough light available
to induce variations of 4\% in polarization in the total light.

A final possibility is that a compact asymmetric structure much closer
to the star is responsible for the variable polarization. As mentioned
earlier, the SED of the object leaves no room for a significant, hot
dusty component. A caveat is of course that the near-infrared
photometry of the object is limited, so any sudden, short-lived events
could have been easily missed. If the ISP value is, as the data
suggest, in the upper half of the {\it QU} diagram, the major
polarization activity occurs when the object occupies the lower half
of the QU diagram. A prediction would then be that during such
periods, excess radiation in the near-infrared due to hot dust would
be present. NIR data of HD 179821 is very sparse, but the last set of
{\it J, K} photometry (from 2MASS), was taken during Melikian's
observing period. These data follow the gradual faintening of the star
in the NIR and do not indicate any excess emission. In addition, the
optical photometry appears steady throughout.

For completeness, we do note that \cite{Oudmaijer_etal:1995} found
variable first-overtone CO emission at 2.3 $\mu$m towards this
object. This emission typically arises in hot, dense material within
the dust sublimation radius. Oudmaijer et al. interpreted this as
variable mass loss. If the CO emission originates from a geometry that
is not spherically symmetric, it could be associated with ejections of
dense gas similar to, but shorter than, for example the asymmetric
mass loss event observed towards HD 45677 \citep{Patel_etal:2006}. The
resulting situation could give rise to Rayleigh scattering and a net
polarization of light. However, whether this would give rise to the
large polarization variations is not clear. We do point out that CO
emission from the yellow hypergiant $\rho$ Cas could also be explained
by shocks due to pulsations in the stellar photosphere
\citep{Gorlova_etal:2006}. This brings us to the central star as the
origin of the variations.

\paragraph{An asymmetric star illuminating the dust shell}

\smallskip

Above, we considered the case
of a central source surrounded by asymmetric scattering media. The
asymmetries involved result in a net polarization of the total
light. However, we tacitly assumed the central object to be either a
point source or an isotropically emitting sphere.

It is well known that when a central source illuminates a spherically
symmetric envelope, any asymmetries in the {\it central source} can
induce a net polarization.  As for example reviewed by
\cite{Boyle_etal:1986}, who find that a photosphere with a nonuniform
surface brightness can result in net polarization when its radiation
is scattered off circumstellar dust.  Other examples are changes found
in the polarization spectra of evolved stars where it has been
inferred that different molecular bands cover different parts of the
stellar photosphere \citep{Bieging_etal:2006}.  The polarization
changes over molecular bands can be explained by both Rayleigh
scattering by particles in clouds high in the atmosphere and
scattering due to circumstellar dust \citep{Raveendran:1991}. In all
these cases the break from symmetry is not due to the circumstellar
material, but due to the asymmetric illuminating source.

The extensive photometric observations and period analysis of
\cite{LeCoroller_etal:2003} indicate that HD~179821 is a non-radial
pulsator (NRP).
It is not yet clear whether the shape of the star changes altogether,
as we lack the spectral data (Le Coroller et al. quote unpublished
data implying that the velocity variations are modest), or due to
changes in surface temperature, the situation Le Coroller et
al. prefer. In either case, the circumstellar material will be
irradiated by an asymmetric source. Then, the polarization changes can be
expected to occur around a preferred axis, as, apart from rotation,
the pulsation nodes are not expected to stochastically change position
on the star.

In order to investigate whether it is plausible that a star with an
aspherical shape can yield observable polarization, we performed some
test calculations of a star covered by large spots, surrounded by a
spherically symmetric dust shell.  The model set-up consists of a star
with a spot. In the situation where one (bright) spot is present, the
polarization will increase with viewing angle. That is to say, if the
spot is located in the line of sight, we will see its light directly
and not much net polarization will be measured. If on the other hand
the system is inclined, less direct light will be seen and the
polarized light contribution increases, hence a larger polarization is
observed. We should also note that unless the scattering occurs very
close to the star (at a few stellar radii), the polarization
percentage is independent of the radius of the shell. This is due to
the constant geometry of the situation.

For the simple case where all the stellar flux is radiated from
one spot covering 10 per cent of the stellar surface, we find that
polarizations up to 3 per cent are readily produced when the star is
mildly inclined. For smaller surface areas, the polarization can be
even larger, exceeding 10 per cent. This result confirms earlier
studies which used less extreme parameters \citep{Raveendran:1991,
AlMalki_etal:1999}.  Of course this is an idealized situation, but it
demonstrates that net polarizations of order per cent are certainly
possible. Indeed, from an observational point of view, polarization
changes of around 1 per cent have been observed towards Betelgeuse
\citep{hayes_1984}. These changes were due to variable hotspots on the
stellar surface.  Therefore, polarization changes can indeed be of
order few per cent if the star itself is anisotropic.

One could argue why the polarization variability is not modulated by
the 140 or 200 day period. Perhaps it is on the shorter term, but we
lack the data to properly investigate this. However, where we could
match photometric and polarimetric data, on shorter timescales it
appears that the polarization values closely follow the photometry.
It is therefore possible that the polarization is variable at similar
timescales to the pulsations. A further question is why such long term
changes are observed. We note that the mean photometry shows long term
changes, probably tracing the global evolution of the photosphere,
which would have an effect on the non-radial pulsations.
Significantly, the preferred plane would not necessarily change, and
this is found.

A critical test of the scenarios discussed above is simultaneous
photo-polarimetric monitoring of HD~179821. If non-radial pulsations
are responsible for the observed polarization, both observables should
be strongly correlated. If, on the other hand, episodic mass ejections
are the main cause of the variations, we would expect enhanced
near-infrared emission due to hot dust during significant polarization
changes.

\section{Conclusion}

We have presented spectropolarimetry and long term photo-polarimetry
for two post-Red Supergiants. A strong depolarization across the
H$\alpha$ emission line is found for IRC~+10420, suggesting an
electron-scattering region that is not circularly symmetric,
confirming the results of \cite{Davies_etal:2007}, who found such
evidence from their integral field spectroscopy of the object. The
time evolution indicates that the source has increased in temperature
until at least 1995, after which the {\it J} band photometry indicates
this may have levelled off.  If the temperature increase of the object
has indeed halted, an increase in mass loss or mass infall rate can be
responsible for the observed H$\alpha$ emission and polarization.

HD~179821, a cooler object, has less H$\alpha$ emission, and no
depolarization across the H$\alpha$ line could be detected. The
photometry implies that this evolved object is also undergoing a
change in temperature. Hitherto, temperature determinations of the
object were inconclusive, but if the star is a G supergiant, then the
observed change in photometry suggests it has become earlier by one or
two subclasses. Strong changes at the 5 per cent level in
polarization over the past 15 years have been detected. During the
same time, the optical photometry has only varied by at most 0.2
magnitudes. The most obvious explanations for this observation such as
changes in either electron or dust scattering can be ruled out.  A
complication is that the polarization is not correlated with any other
observable. The possibility that the star undergoes irregular,
asymmetric, mass ejections is discussed. However, there is little
evidence for the occurrence of such events. Instead, it is proposed
that the star itself is asymmetric and that its anisotropic radiation,
which is scattered off the circumstellar dust results in the net
polarization observed.  


\subsection*{Acknowledgments}

We wish to thank the referee, Prof. Kenneth H. Nordsieck, for his
detailed reading of the paper and constructive comments. We thank
Willem-Jan de Wit for reading the manuscript.  Based on observations
conducted on the WHT, the NOT and the CST on the Canary islands, the
AAT in Australia, and the TSAO in Kazakhstan. RDO is grateful for the
support from the Leverhulme Trust for awarding a Research
Fellowship. JCM and MP and acknowledge PhD student funding from
PPARC/STFC and the University of London respectively. We are grateful
to Joni Johnson for allowing us to use her HPOL data on HD~179821.

\bibliography{mnemonic,Patel_specpolref}
\bibliographystyle{mn2e}
\bsp
\label{last page}
\end{document}